\documentclass[pre,10pt,amsmath,amssymb]{revtex4}
\usepackage{graphicx}

\usepackage{amsmath,amsfonts,amssymb,bm}

\begin{document}

%\draft
\title{Anomalous diffusion for a correlated process with long jumps}

\author
{Tomasz Srokowski}

\affiliation{
 Institute of Nuclear Physics, Polish Academy of Sciences, PL -- 31-342
Krak\'ow,
Poland }

\date{\today}

\begin{abstract}
We discuss diffusion properties of a dynamical system, which is characterised by 
long-tail distributions and finite correlations. The particle velocity has 
the stable L\'evy distribution; it is assumed as a jumping process 
(the kangaroo process) with a variable jumping rate. Both the exponential and 
the algebraic form of the covariance -- defined for the truncated distribution -- 
are considered. It is demonstrated by numerical calculations that the stationary 
solution of the master equation for the case of power-law correlations decays 
with time, but a simple modification of the process makes the tails stable. 
The main result of the paper is a finding that -- in contrast to the velocity 
fluctuations -- the position variance may be finite. It rises with time faster 
than linearly: the diffusion is anomalously enhanced. On the other hand, 
a process which follows from a superposition of the Ornstein-Uhlenbeck-L\'evy processes 
always leads to position distributions with a divergent variance which means 
accelerated diffusion.

\end{abstract}

%\pacs{02.50.Ey,05.40.Ca,05.40.Fb}

\maketitle


\section{Introduction}

Stochastic force in the dynamical description of a physical system is 
usually assumed as a white, Gaussian noise. That approximation may be 
sufficient if the correlation time is short, compared to the time-scale 
of the system. However, correlations of the 
stochastic force are important in many physical problems. Slowly decaying 
correlations are present in various phenomena
including chemical reactions in solutions \cite{gro},
ligand migration in biomolecules \cite{han}, atomic diffusion
through a periodic lattice \cite{iga}, spectroscopy \cite{fri1}
and hydrodynamics \cite{tuck,sane}. Long-time memory may also occur 
in climate, physiology, financial markets and earthquake research 
\cite{lenn,podo,podo1}. The paper Ref.\cite{rome} reported that 
the normalised prices of both 
Babylonian and English agricultural commodities exhibited persistent correlations 
of a power-law type over long periods of up to several centuries.
Moreover, long-time correlations emerge in 
effective, low-dimensional stochastic descriptions of complicated 
deterministic systems; they are produced by the procedure 
of fast degrees of freedom removal \cite{med,han1}. 
Long correlations of the velocity are often connected with the anomalous diffusion: 
the position variance rises either slower or faster than linearly 
with time. Examples of anomalous diffusion are frequently 
encountered in disordered media \cite{bou}. It has been demonstrated 
recently that the subdiffusive motion of a polymer is related 
to scaling of the monomer mean square displacement and the velocity 
autocorrelation function is algebraic \cite{weber}. The importance of 
strong memory (long autocorrelations) -- as well as long tails 
of the distribution-- in fluctuations of company profit was stressed 
in Ref. \cite{rom}: simple, uncorrelated models are a poor approximation 
to real market phenomena. Systems with finite correlation time of 
the stochastic force may be 
described by the integro-differential Langevin equation. If the memory 
kernel has the same form as the force autocorrelation function, 
fluctuation-dissipation relations are satisfied \cite{kubo}. 

Anomalous transport can be described by jumping processes. 
Continuous time random walk theory (CTRW) \cite{met}, in particular, predicts 
subdiffusion which results from the non-Poissonian form of the 
waiting-time distribution and makes the process non-Markovian. 
That property is mathematically expressed as a fractional time-derivative 
in the Fokker-Planck equation. Anomalous diffusion, both super- and 
subdiffusion, is predicted by a Markovian version of CTRW if one allows 
for a variable jumping rate \cite{kam06}. Besides the waiting-time distribution, 
CTRW involves the jumping-size distribution. Due to the central limit theorem, 
stable distributions are distinguished. They can have either the form of 
a Gaussian or a general stable L\'evy distribution, which is 
characterised by long tails $\sim|x|^{-1-\alpha}$, where $0<\alpha<2$ 
is the stability index (L\'evy flights) \cite{shl,shl1}. 
Then, the Fokker-Planck equation contains the fractional derivative in respect 
to the process variable. The above asymptotics implies that variance, as well as all higher 
moments, must be divergent for any time (the accelerated diffusion). This property 
means that infinite jumps are performed during a finite 
time; for that reason L\'evy flights are unphysical in dynamical problems. 
The difficulty can be avoided if one introduces the L\'evy walk for which the jump size is 
strictly related to the time interval \cite{met}. Those systems can exhibit 
anomalous diffusion. 
This is the case for the transport of two-level atoms in optical molasses 
derived from counterpropagating laser beams \cite{mark}. 
The atom which stays for a long time in the potential well may 
accumulate enough energy, due to spontaneous emissions, to jump above 
the barrier and perform a long flight. If the depth of the optical potential 
is small enough, both the momentum autocorrelation function and 
the position variance obey the power-law dependence; diffusion may be 
anomalous. One can expect that for the Gaussian processes the system 
is close to the thermal equilibrium; the detailed balance is then satisfied. 
It has been recently demonstrated that CTRW with a constant 
driving force satisfies the Einstein relation for an arbitrary 
memory kernel \cite{kwok}. 

A jumping process, which is well suited to modelling long-time correlations, 
is called the 'kangaroo process' (KP) \cite{fri}. It is defined by the Poissonian 
waiting-time distribution with a variable rate. Instead of the jump-size 
distribution, KP assumes the distribution of the process value just after the jump. 
As a result, it cannot be described by a differential -- either 
second order or fractional -- Fokker-Planck equation, and then it is well 
suited as a model of non-diffusive processes with strong collisions. 
KP is a natural formalism to handle 
the fully developed boundary-layer turbulence since a local description 
is inadequate in this case \cite{dek}; the velocity profile for the Newtonian 
shear flow is governed by the integro-differential equation of KP. Moreover, 
a stochastic description of the hyperfine structure of spectral lines is 
possible by KP since correlations of the radiation emission/absorption 
operator can be easily taken into account \cite{blume1}. For the same 
reason, KP was applied to describe the stochastic Stark broadening, which 
involves correlations $\sim1/t$ \cite{fri1}. It can also serve as a model 
of coloured noises in dynamical problems \cite{doe,plo,kos}. 

Considerations involving the correlation problem require some care for the case 
$\alpha<2$ since the covariance function is infinite. 
One can introduce a simple modification of the process by cutting the 
distribution at some large value. Such a truncation procedure is realistic 
since all physical systems are finite. For example, truncation of the velocity 
distribution emerges in the natural way in optical molasses since the Doppler 
force becomes large at high momentum \cite{mark}. Processes with truncated 
distributions have finite (co$-$)variance, besides that their properties do not differ 
from those for the stable distributions \cite{tou,mant}. 

The aim of this paper is to demonstrate that diffusion -- in the presence of the L\'evy 
flights -- qualitatively depends on both the correlation form and specific choice of the noise. 
We consider two models of the particle velocity: a jumping process (KP) and the 
Ornstein-Uhlenbeck-L\'evy (OUL) process. The paper is organized as follows. In Sec.II KP 
is defined; its properties and stationary solutions are discussed. 
The distributions of the position and diffusion properties of the system 
are discussed in Sec.III. In Sec.IV the results are compared with the predictions 
of a model which is based on the OUL process.

\section{The jumping process} 

The process is defined in terms of two probability distributions $Q(\xi)$ and $P_P(\tau)$. 
$Q(\xi)$ determines the step-wise process value just after a jump. More precisely, 
the probability that one jump occurred in an infinitesimal time interval $(0,\Delta t)$ 
is given by $\nu(\xi)\Delta t$, where $\nu(\xi)$ denotes a given, variable jumping rate. 
Therefore, the waiting-time density distribution, $P_P(\tau)$, is Poissonian, 
\begin{equation}
\label{poi}
P_P(\tau)=\nu(\xi)\hbox{e}^{-\nu(\xi)\tau}; 
\end{equation}
it determines the probability that the jump occurs for the first time in the interval 
$(\tau,\tau+\Delta t)$. Immediately after the jump, the probability density 
of $\xi$ becomes $Q(\xi)$ and the process value 
remains constant until the next jump. The process is stationary and Markovian, and 
it can be described by the infinitesimal transition probability $\xi'\rightarrow \xi$: 
  \begin{equation}
  \label{trkp}
p_{tr}(\xi,\Delta t|\xi',0)=[1-\nu(\xi') \Delta t]
  \delta(\xi'-\xi)+\nu(\xi') \Delta t Q(\xi).
  \end{equation}
Since the time interval $\Delta t$ is small, only one jump is taken into account 
in the above expression. The first term in Eq.(\ref{trkp}) corresponds to the event 
that no jump occurred. From Eq.(\ref{trkp}), the master equation directly follows: 
  \begin{equation}
  \label{fpkp}
  \frac{\partial}{\partial t}p(\xi,t) = -\nu(\xi)p(\xi,t) + 
 Q(\xi)\int_{-\infty}^\infty \nu (\xi') p(\xi',t) d\xi'.
  \end{equation}
The stationary solution of Eq.(\ref{fpkp}) is given by
\begin{equation}   
\label{psta}
P_1(\xi)=\frac{Q(\xi)\langle \nu(\xi)\rangle}{\nu(\xi)} 
=\frac{Q(\xi)/\nu(\xi)} {\int_{-\infty}^\infty Q(\xi')/\nu(\xi') d\xi'},
\end{equation}
where the frequency $\nu(\xi)$ is averaged over $P_1(\xi)$. Existence of the integral in 
Eq.(\ref{psta}) imposes appropriate requirements on the functions $Q(\xi)$ and $\nu(\xi)$. 
However, $P_1(\xi)$ is not the only stationary solution of Eq.(\ref{fpkp}): 
if $\lim_{\xi\to a}\nu(\xi)/(\xi-a)$ is finite for a constant $a$, a delta function 
also constitutes a solution. Then, for $a=0$, Eq.(\ref{fpkp}) implies 
\begin{equation}
\label{pnsta}
P_2(\xi)=\delta(\xi). 
\end{equation}
Therefore, the master equation (\ref{fpkp}) can have the regular solution  
(Eq.(\ref{psta})), the singular one (Eq.(\ref{pnsta})) or both, 
according to the specific choice of the functions $Q(\xi)$ and $\nu(\xi)$. 

The process is characterised by a broad spectrum of possible covariance 
functions. We define that quantity in the stationary limit by  
a joint probability $P$ as 
\begin{equation}
\label{dcov}
{\cal C}(\tau)=\langle \xi(t)\xi(t+\tau)\rangle/\langle \xi(t)^2\rangle=
\int\int \xi\xi'P(\xi,t+\tau;\xi',t)d\xi d\xi'
/\langle \xi(t)^2\rangle=\int\int \xi\xi'P(\xi,t+\tau|\xi',t)P_1(\xi')d\xi d\xi'
/\langle \xi(t)^2\rangle,
 \end{equation}
where we assume that all the above quantities exist and are finite. 
${\cal C}(t)$ can be evaluated by summing up over all possible sequences 
of jumps \cite{kam03}. If both $P_1(\xi)$ and $\nu(\xi)$ are even functions, 
Eq.(\ref{dcov}) takes a simple form 
\begin{equation}
\label{corin}
{\cal C}(t)=2\int_0^\infty \xi^2 \exp(-\nu(\xi)t)P_1(\xi)d\xi/\langle \xi^2\rangle.
 \end{equation}
The integral in Eq.(\ref{corin}) can be expressed in the form 
$\int_{\nu(0)}^\infty \xi^2 \exp(-\nu(\xi)t)P_1(\xi)(d\xi/d\nu) d\nu$ which is essentially 
a Laplace transform. We want to construct the jumping process for a given  
covariance by an appriopriate choice of the (even) function $\nu(\xi)$. This can be achieved 
by inverting the Laplace transform from ${\cal C}(t)$; then $\nu(\xi)$ follows 
from a simple differential equation 
\begin{equation}
\label{f12}
\frac{d\nu}{d\xi} = 2 \xi^2 P_1(\xi)/{\widetilde{{\cal C}}}(\nu),
\end{equation}
where ${\widetilde{C}}(\nu)$ denotes the inverse Laplace transform of ${\cal C}(t)$.
$\nu(x)$, calculated from Eq.(\ref{f12}), defines the required process as soon as 
$Q(x)$ is an even function. In particular, for the power-law covariance 
\begin{equation}
\label{corp}
{\cal C}(t)\sim 2\theta\Gamma(\theta)t^{-\theta}~~~~~~~~~(\theta>0),
\end{equation}
Eq.(\ref{f12}) yields 
\begin{equation}
\label{nup}
\nu(\xi)=\left[2\int_0^{|\xi|} \xi'^2 P_1(\xi')d\xi'\right]^{1/\theta}.
 \end{equation}

In this paper, we assume the stationary distribution $P_1(\xi)$ in the form 
of the stable, symmetric L\'evy distribution. It is given by the following 
Fourier transform 
\begin{equation}
\label{lev}
P_1(\xi)=\frac{1}{\pi} \int_0^\infty \exp(-D k^\alpha)\cos(k\xi)dk, 
\end{equation}
where the order parameter $\alpha\in(0,2]$. Except for the case $\alpha=2$, which 
corresponds to the normal distribution, $P_1(\xi)$ possesses the algebraic tail 
$\sim |\xi|^{-\alpha-1}$. Then the variance is divergent and the covariance definition, 
Eq.(\ref{corin}), does not directly apply. To overcome that difficulty, one 
introduces modifications to that definition, e.g. the codifference \cite{samr,emb}, which 
can serve as an estimation of the memory loss, or introduces an infinite 
cascade of Poissonian correlation functions \cite{elia}. On the other hand, we can approximate 
the L\'evy distribution by introducing a cut-off at some large value of the argument. 
Those truncated distributions \cite{mant,mant1,podo2,kop,sok,srot} 
are realistic in respect to physical applications, 
and they agree with the exact ones up to arbitrary large $\xi$ \cite{tou}; convergence to 
the normal distribution is very slow. In the following, ${\cal C}(t)$ will be understood 
in the sense of the truncated distributions. 

In the simplest case of constant jumping rate, $\nu(\xi)=\nu_0=$const, the master equation, 
Eq.(\ref{fpkp}), takes the form $\frac{\partial}{\partial t}p(\xi,t)=\nu_0[Q(\xi)-p(\xi,t)]$. 
Its solution with the initial condition $p(\xi,0)=\delta(\xi)$, 
\begin{equation}
\label{roze}
p(\xi,t)=\delta(\xi)\hbox{e}^{-\nu_0t}+Q(\xi)(1-\hbox{e}^{-\nu_0t}), 
\end{equation}
converges with time to the unique stationary distribution $P_1(\xi)=Q(\xi)$. The exponential 
form of the covariance follows directly from Eq.(\ref{corin}). 

Equality of the distributions $Q(\xi)$ and $P_1(\xi)$ does not hold for the variable 
jumping rate $\nu(\xi)$, cf. Eq.(\ref{psta}). We consider the case of the power-law 
covariance, Eq.(\ref{corp}). To ensure the existence of $P_1(\xi)$, 
we impose a condition on the parameters 
$\alpha(\theta+1)>2$; then $\langle \nu(\xi)\rangle<\infty$. On the other hand, 
the singular solution $P_2(\xi)$ (Eq.(\ref{pnsta})) also exists if $\theta<3$ 
because $\nu(\xi)\sim |\xi|^{3/\theta}$ for $|\xi|\ll1$. Which of those two solutions, 
$P_1(\xi)$ or $P_2(\xi)$, is stable? This problem was considered in Ref.\cite{sro00} for 
the constant $P_1(\xi)$ on a finite interval. If one 
approximates the interval length $\tau$ in Eq.(\ref{poi}) by its average, $1/\nu(\xi)$, 
the process is fully determined by the distribution of $\tau$ ($P(\tau)$) 
which, in turn, uniquely follows from $P_1(\xi)$. However, we are interested in the process 
value at a given time $t$, i.e. the last interval in the evolution must contain $t$. 
That requirement imposes a bias, namely longer intervals are more probable. 
As a result, the effective $P(\tau)$ is modified: it acquires a time dependence and 
the effective variance dwindles with time to zero. Therefore, the solution $P_1(\xi)$ 
appears unstable, and it decays to the delta function. In the following, we will 
demonstrate by a numerical analysis that $P_1(\xi)$ in the L\'evy form 
is also unstable. 
\begin{figure}[tbp]
\includegraphics[width=8.5cm]{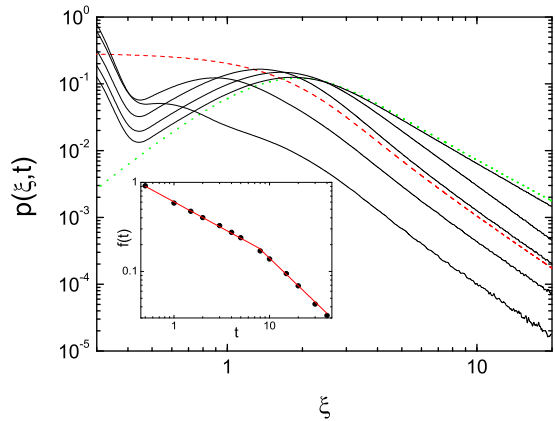}
\caption{(Colour online) Time evolution of the density distribution $p(\xi,t)$ 
for algebraic ${\cal C}(t)$. Solid lines correspond to the following times: 
0.05, 0.5, 2, 10 and 40 (from top to bottom at large $\xi$). The distribution 
$P_1(\xi)$ is marked by the red dashed line and $Q(\xi)$ by green dots. 
Inset: relative decay rate of $p(\xi,t)$ (dots) and estimation of that rate 
by the function $f(t)$, according to Eq.(\ref{fodt}) (solid line).}
\end{figure}

Simulation of the stochastic trajectory requires sampling of the process value 
$\xi$ from the distribution $Q(\xi)$, which is given by 
Eq.(\ref{psta}). Details of that procedure -- for the L\'evy distributed $P_1(\xi)$ 
and the power-law covariance -- are presented in Appendix. The procedure is relatively 
simple if $\theta=1$ and $\alpha=3/2$. In the following, we restrict the numerical 
simulations to that case. Consecutive time intervals are given by the distribution 
(\ref{poi}). The time evolution 
of the density distribution $p(\xi,t)$ is presented in Fig.1, where the initial condition 
$p(\xi,0)=Q(\xi)$ was assumed. The tail quickly converges to the L\'evy shape 
$|\xi|^{-1-\alpha}$ but its relative strength falls with time. 
The rate of the decay can be estimated 
by a function which consists of two power-law segments; it is given by 
\begin{eqnarray}
\label{fodt}
f(t)=
\left\{
\begin{array}{ll}
c_1t^{-\beta_1} &\mbox{for $t\le7.5$} \\
c_2 t^{-\beta_2} &\mbox{for $t>7.5$},
\end{array}
\right.
\end{eqnarray}
where $\beta_1=0.584$, $\beta_2=1.03$, $c_1=0.586$ and $c_2=1.46$. 
Decay of the tail is compensated by an increase of the probability density near $\xi=0$: 
we observe convergence to $P_2(\xi)$. 

The nonsingular solution becomes stable for large $|\xi|$ after an appropriate 
variable transformation, $\xi'=g(t)\xi$. The function $g(t)$ follows from the 
requirement that the tail of the new distribution, 
$P'_1(\xi')\sim|\xi'|^{-\alpha-1}$, is time-independent. Using the identity 
$P'_1(\xi')=P_1(\xi)|d\xi/d\xi'|$, where $P_1(\xi)\sim f(t)|\xi|^{-\alpha-1}$, yields 
\begin{eqnarray}
\label{godt}
g(t)=
\left\{
\begin{array}{ll}
c'_1t^{\beta'_1} &\mbox{for $t\le7.5$} \\
c'_2 t^{\beta'_2} &\mbox{for $t>7.5$},
\end{array}
\right.
\end{eqnarray}
where $\beta'_i=\beta_i/\alpha$ and $c'_i=c_i^{-1/\alpha}$. Numerical calculations, 
presented in Fig.2, confirm that the tail of the limit distribution is stable and 
coincides with the L\'evy one. The decay of the 
distribution $P_1(\xi)$ with time accelerates the decline of the covariance (\ref{dcov}): 
${\cal C}_{eff}^{(1)}(\tau)=f(\tau){\cal C}(\tau)$. On the other hand, 
${\cal C}_{eff}^{(2)}(\tau)=\langle \xi'(t)\xi'(t+\tau)\rangle/\langle \xi'(t)^2\rangle=
g(\tau){\cal C}(\tau)$. Both covariance functions are presented in Fig.2; they follow 
the dependence $\tau^{-1-\beta_1}$ and $\tau^{-1-\beta_1+\beta'_1}$, respectively. 
\begin{figure}[tbp]
\includegraphics[width=8.5cm]{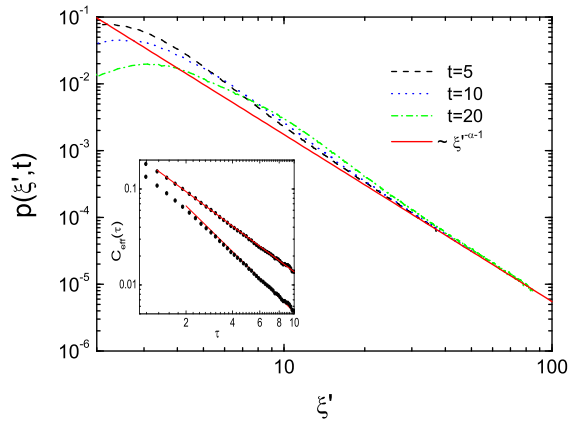}
\caption{(Colour online) Distribution of KP values as a function of the transformed 
variable $\xi'=g(t)\xi$. Inset: effective autocorrelation functions 
$C_{eff}^{(i)}(\tau)$, calculated from the definition at $t=1$. The distribution $Q(\xi)$ 
was truncated at $\xi=10^4$. The solid lines have the slopes 1.584 and 1.195.}
\end{figure}

\section{Diffusion}

Solution of the Fokker-Planck equation for the standard theory of the 
Brownian motion predicts that the position variance rises linearly with 
time. Deviations from that pattern may be caused by nonhomogeneity of the 
medium \cite{bou,kam06}, the presence of regular structures in the phase space 
\cite{gei} or memory in the system \cite{met}. In general, we observe  
the subdiffusion and superdiffusion (the enhanced diffusion), when the variance 
rises slower and faster than linearly with time, respectively. Moreover, 
the accelerated diffusion corresponds to the case of infinite variance 
for any time. If the process is 
stationary and the velocity autocorrelation function ${\cal C}(t)$ exists, 
the time-dependence of the variance is directly related to ${\cal C}(t)$ by 
a simple identity 
\begin{equation}
\label{x2}
\langle x^2(t)\rangle=2\int_0^t(t-\tau){\cal C}(\tau)d\tau, 
\end{equation}
where the average is performed over the stationary distribution. The above 
formula means that slow decrease of ${\cal C}(t)$ implies fast diffusion. 
The non-Markovian CTRW with the Gaussian jump-size distribution predicts 
subdiffusion which results from long waiting times inside the traps. 
Anomalous diffusion may originate not only from temporal characteristics 
of the system but also from the spatial structure of the medium. 
It is the case for transport on fractals \cite{hen} and for the Markovian 
CTRW with position-dependent jumping rate \cite{kam04}; all kinds of diffusion 
are present in those systems \cite{kam06}. The stochastic, additive driving force 
in the Langevin equation with the stable and non-Gaussian distribution leads to the 
accelerated diffusion. The same conclusion holds for CTRW with the stable 
jump-size distribution but introducing a power-law truncation may result 
in heavy tails with convergent variance \cite{sok}.

We look for the distribution $p_x(x,t)$, where the variable $x$ is given by the equation 
\begin{equation}
\label{lanx}
\dot x(t)=v(t)
\end{equation}
and $v(t)$ follows from the jumping process. In the case $\nu(\xi)=$const, which corresponds to 
the exponential covariance, we assume $v(t)=\xi(t)$. The distributions $p_x(x,t)$ were calculated 
from individual trajectories $x(t)$; consecutive process values were sampled from the L\'evy 
distribution by a standard procedure \cite{wer}. Fig.3 presents the results as a function of the 
L\'evy parameter $\alpha$. The tails of all distributions for $\alpha<2$ have the shape 
$|x|^{-\alpha-1}$ which implies infinite variance and accelerated diffusion. The apparent 
width of $p_x(x,t)$ diminishes with $\alpha$ and rises with time. 
\begin{figure}[tbp]
\includegraphics[width=8.5cm]{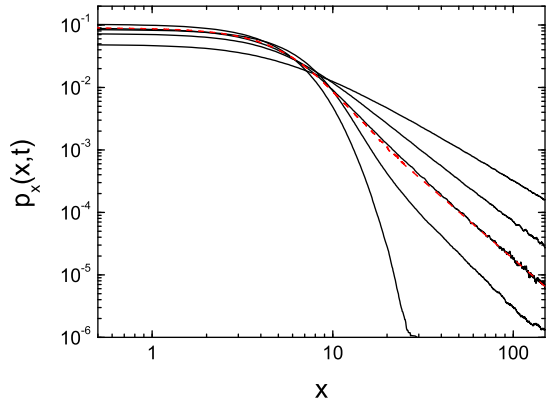}
\caption{(Colour online) Probability density distributions of the variable $x$ for the 
case of the exponential covariance at the time $t=5$. The curves correspond to 
$\alpha=0.8,~1.2,~1.5,~1.8$ and 2 (from right to left); $\nu_0=1$. The red dashed line results from 
inversion of the characteristic function (\ref{sol0}) for $\alpha=1.5$ and $\gamma=1/\alpha$.}
\end{figure}

The case of variable jumping rate involves a coupling between the process values $\xi$ and 
the waiting times $\tau$, cf. Eq.(\ref{nup}). Since a large $\xi$ corresponds to a small 
time interval $\tau$, its influence on the particle displacement is limited, and the 
resulting variance may actually be finite. The condition for the convergence of the variance can be 
estimated in the following way. We assume $v(t)=\xi(t)$ and take into account only the tail 
of the L\'evy distribution, $\sim\xi^{-\alpha-1}$. Then Eq.(\ref{nup}) yields 
$\nu(\xi)\sim\xi^{(2-\alpha)/\theta}$. Approximating the time interval by the average of the 
distribution $P_P(\tau)$, $\tau\approx\langle\tau\rangle=1/\nu$, we obtain the trajectory 
$x(t)$ in the form of a sum of mutually independent variables with the same distribution 
\begin{equation}
\label{xroz}
x(t)=\sum_{i=1}^n \xi_i^{1+(\alpha-2)/\theta},
\end{equation}
where the $n$'th interval contains the time $t$. The distribution of the variable 
$\eta=\xi_i^{1+(\alpha-2)/\theta}$, $P_\eta(\eta)$, is given by 
$P_\eta(\eta)=P_1[\xi(\eta)]|d\xi/d\eta|\sim\eta^{-1-\alpha\theta/(\alpha+\theta-2)}$. 
The variance of the distribution $P_\eta(\eta)$ is finite if 
\begin{equation}
\label{waru}
\frac{\alpha\theta}{\alpha+\theta-2}>2. 
\end{equation}
If the condition (\ref{waru}) is satisfied, the variable $x$ has finite variance 
as well and its distribution converges with time ($n\to\infty$) to the normal 
distribution, though that convergence may be very slow. The variance is always finite 
for the case $\theta=1$, whereas it is always infinite if $\theta\ge2$. 
Therefore, a dynamical system in which the velocity 
is given by KP with the algebraic covariance can be characterised by the finite position 
variance, although the L\'evy stable distribution is assumed. 
This conclusion is the main result of the paper. 
\begin{figure}[tbp]
\includegraphics[width=8.5cm]{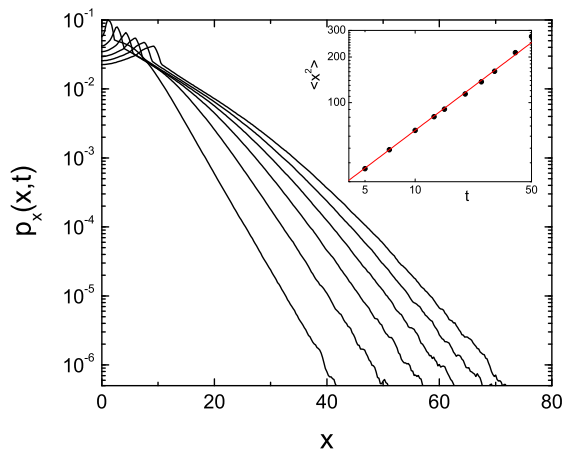}
\caption{(Colour online) Time evolution of the density distribution $p_x(x,t)$ 
for algebraic ${\cal C}(t)$ and $v(t)=\xi(t)$. The curves correspond to the following times: 
5, 10, 15, 20, 25 and 30 (from left to right). Inset: variance as a function of time; 
the straight line marks the dependence $t^{0.83}$.}
\end{figure}

The numerical results for the case $v(t)=\xi(t)$, presented in Fig.4, show that the tails of 
the probability distributions are exponential. The variance rises with time as $t^{0.83}$ 
up to $t=50$, which dependence indicates subdiffusion.  
The case of stationary tails, $v(t)=g(t)\xi(t)$, is presented in Fig.5. The results 
are qualitatively similar to the preceding case but the distributions are broader, and 
they expand faster with time. The shape of the tails is exponential, 
$\sim\exp(-\gamma |x|)$, and the parameter $\gamma$ rises with time. 
The time dependence of the variance is governed by 
the relation $t^{1.65}$ up to $t=40$, whereas for larger times the growth becomes 
even faster. Anyway, the dependence is stronger than linear. 

\begin{figure}[tbp]
\includegraphics[width=8.5cm]{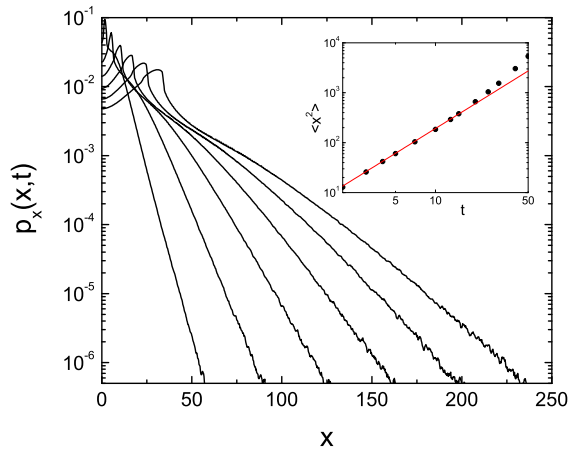}
\caption{(Colour online) The same as Fig.4 but with $v(t)=g(t)\xi(t)$. 
The straight line in the inset marks the dependence $t^{1.65}$.}
\end{figure}

The Langevin equation with uncorrelated noise may predict 
the finite variance if that noise is multiplicative. That problem was 
discussed in Ref.\cite{sro09} for the algebraic multiplicative factor 
$|v|^{-\theta/\alpha}$, where $\theta+\alpha>0$. The conclusions depend 
on a particular interpretation of the stochastic integral. In the It\^o 
interpretation, for which the corresponding Fokker-Planck equation contains a variable 
diffusion coefficient, the tails of the distribution are the same as for the input 
noise: $|v|^{-1-\alpha}$. On the other hand, 
the Stratonovich version predicts the tails $|v|^{-1-\alpha-\theta}$ and the 
process for the case without potential is subdiffusive if $\alpha+\theta>2$. 
Heavy tails with finite variance may also emerge when one introduces a deterministic 
force, both linear and nonlinear \cite{sro10,che}.

\section{Comparison with the Ornstein-Uhlenbeck-L\'evy process}

The Ornstein-Uhlenbeck process describes Brownian particle velocity 
in the framework of the standard Brownian motion theory \cite{gar}. 
The process is defined by the Langevin equation
\begin{equation}
\label{la}
\dot v(t)=-\gamma v(t)+\dot\eta(t), 
\end{equation}
where the white noise $\eta(t)$ generates a process with stationary and 
independent increments. For the Gaussian case, the covariance 
assumes the exponential form with the decay rate $\gamma$ and (\ref{la}) 
is the only process with the above properties, according to the Doob 
theorem \cite{vkam}. In the case of the general stable distributions, 
Eq.(\ref{la}) defines the Ornstein-Uhlenbeck-L\'evy process. Its covariance is 
also exponential when understood in a sense of truncated distributions \cite{sro11}. 
The probability distribution of the variable $v(t)$ is given by the characteristic 
function \cite{jes}
\begin{equation}
\label{solou}
{\widetilde p}(k,t)
=\exp\left[-\frac{1}{\alpha\gamma}|k|^\alpha(1-{\mbox e}^{-\alpha\gamma t})\right]; 
\end{equation}
it converges with time to the stationary distribution. The two-dimensional process 
$(x,v)$ is Markovian and the probability density $p(x,v,t)$ determines the position 
distribution after performing integration over velocity \cite{sro11}. Its characteristic 
function reads 
\begin{equation}
\label{sol0}
{\widetilde p}_x(k,t)={\mbox e}^{-\sigma(t)|k|^\alpha}, 
\end{equation}
where 
\begin{equation}
\label{s}
\sigma(t)=\int_0^g\frac{\kappa^\alpha}{1-\kappa}d\kappa 
\end{equation}
and $g=1-{\mbox e}^{-\gamma t}$. 
According to Eq.(\ref{solou}), OUL process has the same time-scale as KP for 
$\nu_0=\gamma\alpha$, both processes can then be compared. In order to invert 
the transform (\ref{sol0}) we utilise the fact that 
$p_x(x,t)$ is the stable L\'evy distribution and it can be expressed as 
the Fox function \cite{sch}; its numerical values follow from series expansions 
both for small and large $x$. 
%In the latter case the expansion reads 
%\begin{equation}
%\label{sze1}
%p_x(x,t)=\frac{1}{\pi\sigma^{1/\alpha}\alpha}\sum_{n=1}^\infty
%\frac{\Gamma(1+\alpha n)}{n!}\sin(\pi\alpha n/2)
%\left(\frac{|x|}{\sigma^{1/\alpha}}\right)^{-\alpha n-1}.
%\end{equation}
A result of the expansions for $\alpha=1.5$ is compared with the corresponding 
distribution for KP in Fig.3; both distributions agree. 

Processes with long tails of the autocovariance can also be constructed 
by OUL, namely as their superposition. Following 
Ref.\cite{elia1}, we define a non-Markovian compound process in the form  
\begin{equation}
\label{mav}
v(t)=\frac{1}{n}\sum_{i=1}^n v_i(t), 
\end{equation}
where the atomic parts are OUL processes, 
\begin{equation}
\label{laat}
\dot v_i(t)=-\gamma_i v_i(t)+\dot\eta(t), 
\end{equation}
and $n\to\infty$. The parameter $\gamma$ in the atomic process is 
a stochastic variable with a given probability distribution $\phi(\gamma)$ 
which satisfies the normalisation condition $\int_0^\infty \phi(\gamma)d\gamma=1$. 
General solution of Eq.(\ref{mav}) and Eq.(\ref{laat}) over the entire real line is given by 
\begin{equation}
\label{mav1}
v(t)=\lim_{n\to\infty}\int_{-\infty}^t \frac{1}{n}\sum_{i=1}^n{\mbox e}^{-\gamma_i(t-t')}\dot\eta(t')dt'. 
\end{equation}
Covariance follows from the formula for the atomic process, 
Eq.(\ref{dcov}), where the joint probability distribution 
has to be conditioned by $\phi(\gamma)$. The final expression involves the Laplace transform: 
\begin{equation}
\label{cou}
{\cal C}(t)=\int_0^\infty {\mbox e}^{-\gamma t}\phi(\gamma)d\gamma. 
\end{equation}
Eq.(\ref{cou}) can be inverted to obtain $\phi(\gamma)$ for a given form of the covariance. 
Two examples of the power-law form are presented in Fig.6; 
${\cal C}(t)$ was calculated by averaging over an ensemble of trajectories 
from Eq.(\ref{laat}). The case ${\cal C}(t)=(1-{\mbox e}^{-\gamma_0 t})/(\gamma_0t)$ 
corresponds to $\phi(\gamma)=1/\gamma_0$ for $\gamma\le\gamma_0$ and zero elsewhere. 
The second example involves a slowly decaying tail, $\sim t^{-1/2}$, which appears 
when we choose $\phi(\gamma)=1/(2\sqrt{\gamma\gamma_0})$ ($\gamma\le\gamma_0$). 
The emergence of long tails in the autocorrelation function results from the 
presence of small values of $\gamma$ in the integral (\ref{cou}). 
\begin{figure}[tbp]
\includegraphics[width=8.5cm]{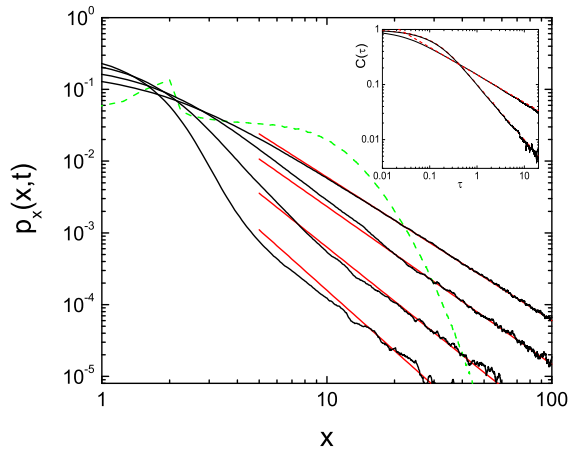}
\caption{(Colour online) Density distributions $p_x(x,t)$ for $v(t)$ which is 
constructed from the OUL processes according to Eq.(\ref{mav}). The solid lines 
correspond to $\alpha=1,~1.2,~1.5$ and 1.8 (from right to left). The solid straight lines 
mark the dependence $x^{-1-\alpha}$ and the dashed green line the result for 
$v(t)$ from KP with $\alpha=1.5$. All distributions were calculated at $t=5$. 
Inset: autocorrelation functions calculated from the trajectory simulations 
with $\phi(\gamma)=1/\gamma_0$ (lower curve) and $\phi(\gamma)=1/(2\sqrt{\gamma\gamma_0})$; 
in both cases $\gamma_0=10$. The lower and upper red dashed lines correspond 
to the functions $(1-{\mbox e}^{-\gamma_0 \tau})/(\gamma_0t)$ and $0.15/\sqrt{\tau}$, 
respectively.}
\end{figure}

Probability distributions of position, $p_x(x,t)$, for $v(t)$ calculated from Eq.(\ref{mav}), 
are presented in Fig.6. All the curves, which correspond to different values of $\alpha$, 
assume the power-law shape of the tails, $|x|^{-\alpha-1}$. Therefore, the variance is divergent. 
Those results are qualitatively different from the distribution for the jumping process 
which is also presented in the figure.

\section{Summary and conclusions}

We have considered a Markovian jumping process, KP, which is characterised by 
long jumps and finite correlation time. Its key property is the dependence of 
the jumping rate on the process value which results, in particular, in 
a variety of correlation functions. The process value distribution is such 
that the corresponding master equation has a stationary solution in the form of 
a L\'evy stable distribution with divergent variance. The correlation functions 
can be properly defined if one introduces a cut-off 
at some large process value. Numerical analysis, 
performed for the stability index $\alpha=3/2$ and the power-law correlation 
function, shows that the stationary solution $P_1(\xi)$ is unstable: 
it decays with time to the delta function. However, the shape of the tails 
converges with time to that of $P_1(\xi)$ and a simple modification 
of the process -- which consists in multiplying the process value by an 
appropriate function of time -- makes the tails stable. 

When the particle velocity is assumed as the L\'evy distributed jumping 
process $\xi(t)$, the transport properties of the system are qualitatively 
dependent on a particular form of its autocorrelation function.
The tails of the position distribution for the exponential 
correlations have the shape $|x|^{-1-\alpha}$, the same as for the velocity, 
indicating infinite variance. It may not be the case for the power-law form 
of ${\cal C}(t)$ since long jumps are penalised by small time intervals. 
It has been demonstrated that if the parameters $\alpha$ and $\theta$ 
satisfy the condition (\ref{waru}), 
the position distribution falls relatively fast making the variance finite. 
Suppressing of long tails of the distribution resembles the 
L\'evy walk approach where the jump length is governed by the time needed to perform 
the jump. The diffusion process for the case of stable tails appears anomalous; 
it is characterised by the variance which rises faster than linearly with time. On 
the other hand, diffusion may be accelerated if condition (\ref{waru}) is not 
satisfied. 

The existence of finite variance of the position distributions -- despite long velocity 
tails -- is not a generic property of power-law correlated processes but 
rather a specific feature of KP. Other processes may exhibit different behaviour; 
we have demonstrated that a superposition of the OUL processes 
is characterised by tails of the form $|x|^{-\alpha-1}$ and diffusion is accelerated, 
similarly to the case of the exponential correlations. Divergence of the variance 
does not violate physical principles for some problems, e.g. the diffusion on 
a polymer chain in chemical space or single molecule spectroscopy \cite{met}. 
However, in dynamical problems with finite particle mass, a finite  
propagation velocity is required. From that point of view, KP offers a reasonable 
alternative to OUL as a model of the noise.

\section*{APPENDIX}

\setcounter{equation}{0}
\renewcommand{\theequation}{A\arabic{equation}} 

Random numbers distributed according to $Q(\xi)$ are generated 
by inversion of the distribution function $\Phi(\xi)$. The following algorithm 
is restricted to the case $\alpha=1.5$ and $\theta=1$. At first, we numerically evaluate 
$\nu(\xi)$ from Eq.(\ref{nup}) for consecutive values of $\xi$ up to $\xi_{max}$, 
with a small step $\Delta \xi$, and store the results. $\xi_{max}$ means 
the process value for which the L\'evy asymptotics is already reached: 
$P_1(\xi)\sim \xi^{-2.5}~~(\xi>\xi_{max})$. Therefore, 
we have either $\nu(\xi)=\int_0^\xi \xi'^2P_1(\xi')d\xi'~~(\xi\le \xi_{max})$ or 
$\nu(\xi)=\nu(\xi_{max})+c(\sqrt{\xi}-\sqrt{\xi_{max}})~~(\xi>\xi_{max})$, where 
$c$ is a constant. Then we evaluate $\langle\nu(\xi)\rangle$ and $Q(\xi)$ from 
Eq.(\ref{psta}). We must invert the function 
$\Phi(\xi)=\int_0^\xi \nu(\xi')P_1(\xi')d\xi'/\langle\nu(\xi)\rangle=r$, where 
$r\in(0,0.5)$ is a uniformly distributed random number. $\Phi(\xi)$ is calculated 
numerically and stored if $r\le\Phi(\xi_{max})$. The asymptotics must be treated 
separately. Performing the integral yields the equation
\begin{equation}
\label{A.1}
\Phi(\xi_{max})+c_1\xi^{-3/2}+c_2\xi^{-1}+c_3-r=0,
\end{equation}
which resolves itself to a third-order algebraic equation. 
Finally, the sign is sampled. In the calculations presented in the paper, the constants 
have the following values: $\xi_{max}=20$, $c=1.228$, $c_1=-2.162$, $c_2=-0.725$ 
and $c_3=0.06038$. 

The above algorithm may be generalised to other values of $\alpha$ and $\theta$ 
but in the most cases Eq.(\ref{A.1}) must be solved numerically. 
Moreover, the integral in the expression for $\Phi(\xi)$ $(\xi>\xi_{max})$ 
may not be an elementary function if $\theta\ne 1$.

\end{document}